\documentclass[a4paper,11pt]{article}
\usepackage[left=3.17cm,right=3.17cm,top=2.54cm,
headheight=0.5cm,headsep=0.54cm,bottom=2.54cm,footskip=0.79cm
]{geometry}

\usepackage{amsmath,amssymb}
\usepackage[pdfstartview=FitH]{hyperref}

\begin{document}

\title{A new kind of representations on noncommutative phase space}

\author{Sicong Jing\thanks{Corresponding author.} , Bingsheng Lin\\
\textit{\small{Department of Modern Physics, University of Science
and Technology of China}}\\
\textit{\small{Hefei, Anhui 230026, China}}
}
\date{20 May 2008}
\maketitle
\footnotetext[1]{\textit{E-mail addresses:} sjing@ustc.edu.cn (S. Jing), xylbs@ustc.edu (B. Lin).}

\begin{abstract}
\noindent
We introduce new representations to formulate quantum mechanics on
noncommutative phase space, in which both coordinate-coordinate and
momentum-momentum are noncommutative. These representations
explicitly display entanglement properties between degrees of
freedom of different coordinate and momentum components. To show
their potential applications, we derive explicit expressions of
Wigner function and Wigner operator in the new representations, as
well as solve exactly a two-dimensional harmonic oscillator on the
noncommutative phase plane with both kinetic
coupling and elastic coupling.
\end{abstract}

\section{Introduction}\label{sec1}
As is well known, representations and transformation theories,
founded by Dirac \cite{s1}, play basic and important role in quantum
mechanics. Many quantum mechanics problems were solved cleverly by
working in specific representations. Some representations, such as,
the coordinate, the momentum, the number representation, as well as
the coherent state representation, are often employed in the
literature of ordinary quantum mechanics. In noncommutative quantum
mechanics (NCQM) \cite{s2}-\cite{s4}, because of the noncommutativity of
coordinate-component (and/or momentum-component) operators, there
are no simultaneous eigenstates for these different coordinate (or
momentum) operators, and one can hardly construct a coordinate (or
momentum) representation in the usual sense. However, in order to
formulate quantum mechanics on a noncommutative phase space so that
some dynamic problems can be solved, we do need some appropriate
representations. Here some few words on the noncommutativity of
momentum-component operators are useful. Although in string theory
only the coordinate space exhibits a noncommutative structure, some
authors have studied models in which a noncommutative geometry is
defined on the whole phase space \cite{s5,s6}. Noncommutativity between
momenta arises naturally as a consequence of noncommutativity
between coordinates, as momenta are defined to be the partial
derivatives of the action with respect to the noncommutative
coordinates \cite{s7}.

On the other hand, the usual method to study NCQM is using the
Seiberg-Witten map to change a problem of NCQM into a corresponding
problem of quantum mechanics on the commutative space. In the case
of only the coordinate space is noncommutative, this method is
consistent with the Weyl-Moyal correspondence which amounts to
replacing the usual product in noncommutative space by a star
product in ordinary space. This method, however, does not always
work, for example, when both coordinates and momenta are
noncommutative, i.e., on a noncommutative phase space, although
using the Seiberg-Witten map one can write down a Hamiltonian of
NCQM in terms of the ordinary commutative coordinates and momenta
operators, one has no way to get a well-defined Schr\"{o}dinger
equation which is consistent with the corresponding star product.
Therefore, it is necessary to develop other new method to solve the
quantum mechanical problems on noncommutative phase space.

Noticing that although two coordinate-component (and/or
momentum-component) operators on the noncommutative phase space do
not commute each other, the difference of the two coordinate
operators indeed commute with the sum of the relevant two momentum
operators, thus we can still employ Einstein-Podolsky-Rosen's (EPR)
\cite{s8} idea to construct entangled states on the noncommutative
phase space. It is easily to show that the entangled states with
continuum variables are orthonormal and satisfy completeness
relations, therefore they present new representations for NCQM. The
first bipartite entangled state representation of continuum
variables is constructed by H. Fan and J. R. Klauder \cite{s9} based
on the idea of quantum entanglement initiated by EPR who used
commutative property of two particles' relative coordinate and total
momentum. In this Letter, for the noncommutative phase space, by this
we mean that \textbf{both} the coordinate-coordinate and the
momentum-momentum operators are noncommutative, we construct
continuum entangled state representations and study their some basic
properties including orthonormality and completeness. Besides, we
derive explicit expressions of Wigner function (WF) and Wigner
operator in the new representations. To show the potential
applications of the entangled state representations in NCQM, we
solve exactly the energy level and WFs of a two-dimensional harmonic
oscillator on a noncommutative phase plane with both kinetic
coupling and elastic coupling.

The work is arranged as follows: In section \ref{sec2} we construct the
entangled state representations for the noncommutative phase space
from a set of deformed boson commutation relations. These entangled
states are orthonormal and complete, so that arbitrary state can be
expended as their linear combination. We also evaluate matrix
elements of noncommutative coordinate and momentum operators in
these representations. Section \ref{sec3} is devoted to derive explicit
expressions of the WFs and the Wigner operator in the new
representation. To show the potential role of the new
representation, we study a two-dimensional oscillator on the
noncommutative phase space with both kinetic coupling and elastic
coupling in section \ref{sec4}, and obtain its energy spectrum and WFs
exactly. Some summary and comments are in the last section.

\section{Entangled state representations for NCQM}\label{sec2}
Without loss of generality and for the sake of simplicity, we only
discuss the four-dimensional noncommutative phase space, in which
both coordinate-coordinate and momentum-momentum are noncommutative.
Operators $\hat{x}$, $\hat{y}$, $\hat{p}_{x}$ and $\hat{p}_{y}$
satisfy the following commutation relations
\begin{equation}\label{xp}
[\hat{x},\,\hat{y}]=\emph{i}\,\mu,~~~~~~~~[\hat{p}_x,\,\hat{p}_y]=
\emph{i}\,\nu,~~~~~~~~ [\hat{x},\,\hat{p}_{x}]= [\hat{y},\,
\hat{p}_{y} ]=\emph{i}\,\hbar,
\end{equation}
and other commutators of these operators are vanishing, where $\mu$
and $\nu$ are real nonzero parameters with dimension of
$(length)^2$ and $(momentum)^2$ respectively. Considering the
following operators
\begin{equation}
\hat{R}=\frac{\hat{x}-\hat{y}}{\sqrt{2}},~~~~
\hat{P}=\frac{\hat{p}_{x}+\hat{p}_{y}}{\sqrt{2}},~~~~
\hat{Q}=\frac{\hat{x}+\hat{y}}{\sqrt{2}},~~~~
\hat{K}=\frac{\hat{p}_{x}-\hat{p}_{y}}{\sqrt{2}}.
\end{equation}
obviously one finds that $\hat{R}$ and $\hat{P}$ are commute each
other, as well as $\hat{Q}$ and $\hat{K}$ are commute each other,
respectively. Thus $\hat{R}$ and $\hat{P}$ have simultaneous
eigenstates $|\lambda \rangle $, and $\hat{Q}$ and $\hat{K}$ have
simultaneous eigenstates $|\xi \rangle $. Here $\lambda$ and $\xi$
may be complex numbers, ($\lambda=\lambda_{1}+\emph{i}\,
\lambda_{2}$, and $\xi = \xi_{1}+\emph{i}\,\xi_{2}$), and
$\lambda_{1}$, $\lambda_{2}$, $\xi_{1}$ and $\xi_{2}$ are real
numbers.

In order to get explicit expressions of the eigenstates $|\lambda
\rangle $ and $|\xi \rangle $, we use the following quadrature
decomposition
\begin{eqnarray}\label{xpab}
&&\hat{x}= \sqrt{\frac{\hbar}{2}} \sqrt[4]{\frac{\mu}{\nu}}(\hat{a} +
\hat{a}^\dag ), ~~~~
\hat{p}_x = \frac{1}{\emph{i}}
\sqrt{\frac{\hbar}{2}} \sqrt[4]{\frac{\nu}{\mu}}(\hat{a}-
\hat{a}^\dag),\nonumber\\
&&\hat{y}= \sqrt{\frac{\hbar}{2}}
\sqrt[4]{\frac{\mu}{\nu}}(\hat{b} + \hat{b}^\dag ), ~~~~~
\hat{p}_y =
\frac{1}{\emph{i}} \sqrt{\frac{\hbar}{2}}
\sqrt[4]{\frac{\nu}{\mu}}(\hat{b}- \hat{b}^\dag).
\end{eqnarray}
Obviously, these dimensionless operators $\hat{a}$, $\hat{b}$ and
their Hermitian conjugates satisfy the following commutation
relations
\begin{equation}\label{abcr}
[\hat{a},\,\hat{a}^\dag ]=[\hat{b},\,\hat{b}^\dag
]=1,~~~~[\hat{a},\,\hat{b}]=[\hat{a}^\dag ,\,\hat{b}^\dag
]=0,~~~~[\hat{a},\,\hat{b}^\dag ]= -[\hat{b},\,\hat{a}^\dag ]=
\emph{i}\,\theta,
\end{equation}
where the dimensionless parameter $\theta= \frac{\sqrt{\mu
\nu}}{\hbar}$. The algebraic relations in Eq.(\ref{abcr}) are exactly same in
form as the deformed boson algebra in \cite{s10}. To our knowledge,
this type of deformed boson commutation relations also appeared in
early work by Caves and Schumaker \cite{s11}. Their quadrature-phase
amplitudes satisfy the same commutation relations as Eq.(\ref{abcr}). When
$\theta=0$, the Eq.(\ref{abcr}) reduces to ordinary boson algebra. It is
worth pointing out that such a deformed boson algebraic relations
derived in \cite{s10} were under an assumption of maintaining
Bose-Einstein statistics (this assumption is equivalent to propose a
direct proportionality between the noncommutative parameters $\mu$
and $\nu$). However, recently Bertolami and Rosa argued that there
is no strong argument supporting such a direct proportionality
relation \cite{s12}. Here we obtain Eq.(\ref{abcr}) from Eq.(\ref{xp}) without using
any similar assumption, besides the quadrature decomposition (\ref{xpab}).

In terms of these deformed boson operators, the operators $\hat{R}$,
$\hat{P}$, $\hat{Q}$ and $\hat{K}$ can be expressed as
\begin{eqnarray}
&&\hat{R}=\frac{\sqrt{\hbar}}{2} \sqrt[4]{\frac{\mu}{\nu}} (\hat{a}
+ \hat{a}^\dag - \hat{b} - \hat{b}^\dag ), ~~~~~~\hat{P}=
\frac{\sqrt{\hbar}}{2\emph{i}} \sqrt[4]{\frac{\nu}{\mu}} (\hat{a} -
\hat{a}^\dag + \hat{b} - \hat{b}^\dag ), \nonumber\\
&&\hat{Q}=\frac{\sqrt{\hbar}}{2} \sqrt[4]{\frac{\mu}{\nu}} (\hat{a}
+ \hat{a}^\dag + \hat{b} + \hat{b}^\dag ), ~~~~~~\hat{K}=
\frac{\sqrt{\hbar}}{2\emph{i}} \sqrt[4]{\frac{\nu}{\mu}} (\hat{a} -
\hat{a}^\dag - \hat{b} + \hat{b}^\dag ).
\end{eqnarray}

The simultaneous eigenstate $|\lambda \rangle$ of $\hat{R}$ and
$\hat{P}$ can be written as
\begin{equation}\label{lam}
|\lambda \rangle= \exp{\left(-\frac{|\lambda|^{2}}{2}+ \theta
\,\lambda_1 \lambda_2 + \lambda  \hat{a}^{\dag}-\lambda^{\ast}
\hat{b}^{\dag}+ \frac{1}{1-\theta^2} (\hat{a}^{\dag} \hat{b}^\dag -
\frac{\emph{i}\,\theta}{2} \hat{a}^{\dag 2} +
\frac{\emph{i}\,\theta}{2}\hat{b}^{\dag 2}) \right)} |00 \rangle,
\end{equation}
where $|00 \rangle$ is a two-mode normalized boson vacuum state in
the deformed Fock space satisfying $\hat{a}\,|00 \rangle =0$,
$\hat{b}\,|00 \rangle=0$ and $\langle 00 |00 \rangle =1$. Sometimes
$| \lambda \rangle$ is written as $|\lambda_1, \lambda_2 \rangle$ to
reflect its parameter dependence. Using the deformed boson
commutation relation (\ref{abcr}), one finds
\begin{equation}\label{ablam}
(\hat{a}- \hat{b}^\dag )|\lambda \rangle =(\lambda -\emph{i}\,
\theta \lambda^\ast) |\lambda \rangle, ~~~~~~~~(\hat{b}-
\hat{a}^\dag )|\lambda \rangle = -(\lambda^\ast + \emph{i}\, \theta
\lambda ) |\lambda \rangle,
\end{equation}
which lead to
\begin{equation}\label{rplam}
\hat{R} | \lambda \rangle = \sqrt{\hbar} \sqrt[4]{\frac{\mu}{\nu}} (
\lambda_1 - \theta \lambda_2) | \lambda \rangle, ~~~~~~~~ \hat{P} |
\lambda \rangle = \sqrt{\hbar} \sqrt[4]{\frac{\nu}{\mu}} ( \lambda_2
- \theta \lambda_1) | \lambda \rangle .
\end{equation}
Thus $| \lambda \rangle$ indeed is the simultaneous eigenstate of
the operators $\hat{R}$ and $\hat{P}$.

Now we are at the position to consider the orthonormality of the $|
\lambda \rangle$. From Eq.(\ref{ablam}) one has
\begin{eqnarray}
&&\langle \lambda' |(\hat{a} - \hat{b}^\dag )|\lambda \rangle =
(\lambda - \emph{i}\,\theta \lambda^\ast) \langle \lambda' | \lambda
\rangle = (\lambda' - \emph{i}\, \theta \lambda^{' \ast }) \langle
\lambda' |\lambda \rangle, \nonumber\\
&&\langle \lambda' |(\hat{b} - \hat{a}^\dag )|\lambda \rangle =
(-\lambda^\ast - \emph{i}\,\theta \lambda) \langle \lambda' |
\lambda \rangle = (-\lambda^{' \ast} - \emph{i}\, \theta \lambda') \langle
\lambda' |\lambda \rangle,
\end{eqnarray}
or $(\lambda -\lambda' -\emph{i}\,\theta (\lambda^\ast - \lambda^{'
\ast})) \langle \lambda' |\lambda \rangle =0$ and $(\lambda^\ast
-\lambda^{'\ast} + \emph{i}\,\theta (\lambda - \lambda')) \langle
\lambda' |\lambda \rangle =0$, which imply that
\begin{equation}\label{lamorth}
\langle \lambda' |\lambda \rangle \varpropto \delta^{(2)}
\big(\lambda -\lambda' -\emph{i}\,\theta (\lambda^\ast - \lambda^{'
\ast}) \big) = \delta^{(2)} \big(\lambda^\ast -\lambda^{'\ast} +
\emph{i}\,\theta (\lambda - \lambda') \big),
\end{equation}
where $\delta^{(2)}(z)\equiv \delta(z_1)\delta(z_2)$, $(z=z_1+{\rm i}z_2)$.
Without loss generality, one may regard the noncommutative parameter
$\theta$ as a very small one. Eq.(\ref{lamorth}) means that only when $\lambda -\lambda'
=\emph{i}\,\theta (\lambda^\ast - \lambda^{' \ast})$ as well as
$\lambda^\ast -\lambda^{'\ast} =- \emph{i}\,\theta (\lambda -
\lambda')$ the inner product $\langle \lambda' |\lambda \rangle$ has
nonzero value. Substituting the latter into the former, one has
$\lambda -\lambda' = \theta^2 (\lambda -\lambda')$, which means that
only for the case of $\lambda -\lambda' =0$, $\langle \lambda'
|\lambda \rangle$ does not vanish (please notice here  $\theta^2
\neq 1$). It is also true that only for the case of $\lambda^\ast -
\lambda^{' \ast} =0$, $\langle \lambda' |\lambda \rangle$ does not
vanish. Therefore, we have $\langle \lambda' |\lambda \rangle
\varpropto \delta^{(2)} (\lambda- \lambda')$. In addition,
for the later convenience, we take the proportional coefficient here
as $\pi/\sqrt{1-\theta^2}$, so we have
\begin{equation}\label{lamorth1}
\langle \lambda' | \lambda \rangle = \frac{\pi}{\sqrt{1- \theta^2}}
\delta^{(2)} (\lambda -\lambda').
\end{equation}

Then we show that $|\lambda \rangle$ satisfies the following
complete relation
\begin{equation}\label{lamcpl}
\frac{\sqrt{1-\theta^2}}{\pi} \int d^2 \lambda \, |\lambda \rangle
\langle \lambda |=1,
\end{equation}
where $d^2 \lambda =d\lambda_1 d\lambda_2$. In order to show it, we
use an expression for the vacuum projection operator in the deformed
Fock space
\begin{equation}
|00 \rangle \langle 00| = ~:\exp{\Big(- \frac{1}{1-\theta^2}\big(
\hat{a}^\dag \hat{a} + \hat{b}^\dag \hat{b} - \emph{i}\,\theta
(\hat{a}^\dag \hat{b} - \hat{b}^\dag \hat{a}) \big) \Big)}:~,
\end{equation}
which will reduce to ordinary form when $\theta =0$, where the
notation $:...:$ means take the normal ordering product for the
operators $\hat{a}, \hat{b}$ and their Hermitian conjugate
operators. Substituting Eq.(\ref{lam}) into Eq.(\ref{lamcpl}), one has
\begin{multline}
\frac{\sqrt{1-\theta^2}}{\pi}\int d\lambda_1 d\lambda_2
:\exp \big( -\lambda_1^2 -\lambda_2^2 + 2\theta \lambda_1 \lambda_2
+ \lambda \hat{a}^\dag -\lambda^\ast \hat{b}^\dag + \lambda^\ast
\hat{a} - \lambda \hat{b}\big) \\
~~~~~~~~~~~~~~~~~~~~\times \exp\Big\{\frac{1}{1-\theta^2}
\big( \hat{a}^\dag \hat{b}^\dag +
\hat{a}\hat{b} -\frac{\emph{i}\,\theta}{2}(\hat{a}^{\dag 2} -
\hat{b}^{\dag 2} -\hat{a}^2 + \hat{b}^2) \big) \\
 - \big( \hat{a}^\dag
\hat{a} + \hat{b}^\dag \hat{b} - \emph{i}\,\theta (\hat{a}^\dag
\hat{b} - \hat{b}^\dag \hat{a}) \big)  \Big\} :\, .~~~~~~~~~~~~~
\end{multline}
Now within the normal ordering product, one can treat the operators
as $c$-numbers, so after integrating over $\lambda_1$ and $\lambda_2$,
one gets the identity operator, which concludes the proof of
Eq.(\ref{lamcpl}).

Similarly, we may express the simultaneous eigenstate of $\hat{Q}$
and $\hat{K}$ as
\begin{equation}
|\xi \rangle= \exp{\left(-\frac{|\xi|^{2}}{2}- \theta \,\xi_1 \xi_2
+ \xi  \hat{a}^{\dag}+ \xi^{\ast} \hat{b}^{\dag} -
\frac{1}{1-\theta^2} (\hat{a}^{\dag} \hat{b}^\dag -
\frac{\emph{i}\,\theta}{2} \hat{a}^{\dag 2} +
\frac{\emph{i}\,\theta}{2}\hat{b}^{\dag 2}) \right)} |00 \rangle,
\end{equation}
which satisfies
\begin{equation}\label{qkxi}
\hat{Q} |\xi \rangle = \sqrt{\hbar} \sqrt[4]{\frac{\mu}{\nu}} (\xi_1
+ \theta \xi_2) |\xi \rangle , ~~~~~~~~\hat{K} |\xi \rangle = \sqrt{\hbar}
\sqrt[4]{\frac{\nu}{\mu}} (\xi_2 + \theta \xi_1) |\xi \rangle .
\end{equation}
Also the $|\xi \rangle$ obeys the following orthonormal and complete
relations
\begin{equation}\label{xi}
\langle \xi'|\xi \rangle = \frac{\pi}{\sqrt{1-\theta^2}}
\delta^{(2)} (\xi-\xi'),~~~~~~~~\frac{\sqrt{1-\theta^2}}{\pi} \int
d^2 \xi \, |\xi \rangle \langle \xi | =1, ~~~~~~(d^2 \xi= d\xi_1
d\xi_2).
\end{equation}

Thus the eigenstates $|\lambda \rangle$ and $|\xi \rangle$ form two
representations for quantum mechanics on the noncommutative phase
space, respectively. Since when $\theta =0$ the states $|\lambda
\rangle$ and $|\xi \rangle$ are called the continuum entangled
states, so one may name these representations in the noncommutative
phase space as the entangled state representations. Sometimes
working in the $|\lambda \rangle$ or $|\xi \rangle$ representation
is convenient, so we first need to know the scalar product of
$|\lambda \rangle$ and $|\xi \rangle$. To do this, considering
$[\hat{a},\,\hat{b}]=0$, we introduce simultaneous eigenstates of
$\hat{a}$ and $\hat{b}$, i.e., the two-mode coherent states on the
noncommutative phase space
\begin{equation}
|\alpha,\beta \rangle = \exp{(\alpha \hat{a}^\dag + \beta
\hat{b}^\dag - \alpha^\ast \hat{a} -\beta^\ast \hat{b})} |00
\rangle,
\end{equation}
which satisfies $\hat{a}|\alpha,\beta \rangle= (\alpha +
\emph{i}\,\theta \beta) |\alpha,\beta \rangle $ and
$\hat{b}|\alpha,\beta \rangle= (\beta - \emph{i}\,\theta \alpha)
|\alpha,\beta \rangle$, respectively. The inner product of two such
coherent states is easily to get
\begin{eqnarray}
\langle \alpha',\beta'|\alpha,\beta \rangle
 &=& \exp \Big\{
-\frac{1}{2}\big(|\alpha|^2 +|\beta|^2 +|\alpha'|^2 +|\beta'|^2\big)
 +\alpha'^{\ast} \alpha + \beta'^{\ast} \beta \nonumber\\
&&+ \frac{\emph{i}\,\theta}{2}(\beta^{\ast}\alpha -
\alpha^{\ast}\beta +\beta'^{\ast}\alpha'-\alpha'^{\ast}\beta')
+\emph{i}\,\theta (\alpha'^{\ast}\beta - \beta'^{\ast}\alpha)
\Big\},
\end{eqnarray}
which means that the two-mode coherent states are normalized but not
orthogonal to each other, and besides, they are over-complete. The
corresponding resolution of the identity in the noncommutative phase
space is
\begin{equation}\label{abcor}
(1-\theta^2) \int \frac{d^2 \alpha\,d^2 \beta}{\pi^2} |\alpha,\beta
\rangle\,\langle \alpha,\beta |=1,
\end{equation}
where $d^2 \alpha = d Re\alpha\, d Im \alpha$ and $d^2 \beta =d Re
\beta \,d Im \beta$. Using Eq.(\ref{abcor}) one gets
\begin{eqnarray}\label{lamxi}
\langle \lambda | \xi \rangle
&=& \frac{1- \theta^2}{\pi^2} \int d^2
\alpha d^2 \beta \langle \lambda | \alpha, \beta \rangle \langle
\alpha, \beta | \xi \rangle \nonumber\\
&=&\frac{1}{2} \exp{\big( \emph{i}\,
(\lambda_1 \xi_2 - \lambda_2 \xi_1 ) + \emph{i}\,\theta (\lambda_1
\xi_1 - \lambda_2 \xi_2) \big)}.
\end{eqnarray}

Having the Eq.(\ref{lamxi}), one easily gets all of matrix elements of the
basic operators $\hat{x}$, $\hat{y}$, $\hat{p}_{x}$ and
$\hat{p}_{y}$ on the noncommutative phase space in the entangled
state representation $|\lambda \rangle$. To do this, we only need to
evaluate $\langle \lambda|\hat{Q}|\lambda' \rangle$ and $\langle
\lambda|\hat{K}|\lambda' \rangle$. With the aid of Eqs.(\ref{qkxi}) and
(\ref{xi}), one gets
\begin{eqnarray}
\langle \lambda |\hat{Q}| \lambda' \rangle &=& \frac{\sqrt{1-
\theta^2}}{\pi} \int d^2 \xi \langle \lambda | \hat{Q} |\xi \rangle
\langle \xi |\lambda' \rangle = \emph{i}\,\sqrt{\hbar}
\sqrt[4]{\frac{\mu}{\nu}} \frac{\partial}{\partial
\lambda_2} \delta^{(2)} (\lambda - \lambda'), \nonumber\\
\langle \lambda |\hat{K}| \lambda' \rangle &=& \frac{\sqrt{1-
\theta^2}}{\pi} \int d^2 \xi \langle \lambda | \hat{K} |\xi \rangle
\langle \xi |\lambda' \rangle = -\emph{i}\,\sqrt{\hbar}
\sqrt[4]{\frac{\nu}{\mu}}  \frac{\partial}{\partial \lambda_1}
\delta^{(2)} (\lambda - \lambda').
\end{eqnarray}
Thus in the $|\lambda \rangle$ representation, we have
\begin{eqnarray}\label{lamxyp}
&&\langle \lambda|\hat{x}|\lambda' \rangle =\sqrt{\frac{\hbar}{2}}
\sqrt[4]{\frac{\mu}{\nu}} \big( \lambda_1 - \theta \lambda_2 +
\emph{i}\, \frac{\partial}{\partial \lambda_2} \big) \delta^{(2)}
(\lambda -\lambda' ), \nonumber\\
&&\langle \lambda|\hat{y}|\lambda'
\rangle =\sqrt{\frac{\hbar}{2}} \sqrt[4]{\frac{\mu}{\nu}} \big( -
\lambda_1 + \theta \lambda_2 + \emph{i}\, \frac{\partial}{\partial
\lambda_2} \big) \delta^{(2)}
(\lambda -\lambda'), \nonumber\\
&&\langle \lambda|\hat{p}_x|\lambda' \rangle =\sqrt{\frac{\hbar}{2}}
\sqrt[4]{\frac{\nu}{\mu}} \big( \lambda_2 - \theta \lambda_1 -
\emph{i}\, \frac{\partial}{\partial \lambda_1} \big) \delta^{(2)}
(\lambda -\lambda' ), \nonumber\\
&&\langle \lambda|\hat{p}_y|\lambda' \rangle =\sqrt{\frac{\hbar}{2}}
\sqrt[4]{\frac{\nu}{\mu}} \big( \lambda_2 - \theta \lambda_1 +
\emph{i} \, \frac{\partial}{\partial \lambda_1} \big) \delta^{(2)}
(\lambda -\lambda').
\end{eqnarray}

Similarly, in the $|\xi \rangle$ representation, we have
\begin{eqnarray}
&& \langle \xi|\hat{x}|\xi' \rangle =\sqrt{\frac{\hbar}{2}}
\sqrt[4]{\frac{\mu}{\nu}} \big( \xi_1 + \theta \xi_2 + \emph{i}\,
\frac{\partial}{\partial \xi_2} \big) \delta^{(2)}
(\xi -\xi' ), \nonumber\\
&& \langle \xi|\hat{y}|\xi' \rangle =\sqrt{\frac{\hbar}{2}}
\sqrt[4]{\frac{\mu}{\nu}} \big( \xi_1 + \theta \xi_2 -\emph{i}\,
\frac{\partial}{\partial \xi_2} \big) \delta^{(2)}
(\xi -\xi' ), \nonumber\\
&& \langle \xi|\hat{p}_x|\xi' \rangle =\sqrt{\frac{\hbar}{2}}
\sqrt[4]{\frac{\nu}{\mu}} \big( \xi_2 + \theta \xi_1 - \emph{i}\,
\frac{\partial}{\partial \xi_1} \big) \delta^{(2)}
(\xi -\xi' ), \nonumber\\
&& \langle \xi|\hat{p}_y|\xi' \rangle =\sqrt{\frac{\hbar}{2}}
\sqrt[4]{\frac{\mu}{\nu}} \big( -\xi_2 - \theta \xi_1 -\emph{i}\,
\frac{\partial}{\partial \xi_1} \big) \delta^{(2)} (\xi -\xi' ).
\end{eqnarray}

\section{Wigner operator in the continuum entangled representation}
\label{sec3}
The so-called Wigner operator (or Wigner-Weyl quantizer)
$\hat{\triangle} (x,p)$ is an integral kernel in phase space which
transfers a classical function $h(x,p)$ to a quantum operator
$\hat{H} (\hat{x},\hat{p})$
\begin{equation}
\hat{H}(\hat{x},\hat{p}) = \int dx\, dp \,h(x,p) \hat{\triangle}
(x,p),
\end{equation}
and this quantization prescription is referred to as Weyl
correspondence \cite{s13}. In fact, because of noncommutativity
between different operators in quantum mechanics, the corresponding
quantum operator of a classical function is uncertainty. According
to Weyl and Wigner \cite{s14}, the $\hat{\triangle} (x,p)$ is
\begin{equation}
\hat{\triangle}(x,p)= \frac{1}{(2\pi \hbar)^2} \int du\, dv
\exp{\big( \frac{\emph{i}}{\hbar} u(\hat{p}-p) + \frac{
\emph{i}}{\hbar}v(\hat{x}-x) \big)}.
\end{equation}
It is well-known that in ordinary quantum mechanics, in coordinate
representation, the Wigner operator $\hat{\triangle} (x,p)$ has the
form
\begin{equation}\label{delxp}
\hat{\triangle} (x,p)= \frac{1}{2\pi} \int dv \,e^{-\emph{i}\,vp}
|x- \frac{\hbar v}{2} \rangle \langle x+ \frac{\hbar v}{2} |,
\end{equation}
where $|x \rangle$ is the eigenstate of coordinate operator. Eq.(\ref{delxp})
leads to
\begin{equation}
Tr[\triangle (x,p) \triangle (x',p')]= \frac{1}{2 \pi \hbar} \delta
(x-x') \delta (p-p').
\end{equation}
Therefore, for a given operator $\hat{H} (\hat{x},\hat{p})$, its
classical correspondence is
\begin{equation}
h(x,p)= 2\pi \hbar \,Tr[\hat{\triangle} (x,p)
\hat{H}(\hat{x},\hat{p})],
\end{equation}
and for arbitrary state $|\psi \rangle$, the corresponding Wigner
function is $W(x,p) =\langle \psi |\hat{\triangle} (x,p)|\psi
\rangle$.

In the noncommutative phase space, the Wigner operator is
\begin{equation}
\hat{\triangle}(\mathbf{x},\mathbf{p})= \frac{1}{(2\pi \hbar)^4}
\int d\mathbf{u}\, d\mathbf{v} \exp{\big( \frac{\emph{i}}{\hbar}
\mathbf{u} \cdot (\hat{\mathbf{p}}-\mathbf{p}) + \frac{
\emph{i}}{\hbar} \mathbf{v} \cdot (\hat{ \mathbf{x}}- \mathbf{x})
\big)},
\end{equation}
where $\mathbf{u}=(u_1,u_2)$, $\mathbf{v}=(v_1,v_2)$,
$\mathbf{x}=(x,y)$, $\mathbf{p}=(p_x,p_y)$ and $\hat{\mathbf{x}}$,
$\hat{\mathbf{p}}$ are the corresponding quantum operators
respectively. Now we consider the WF of the entangled state $|
\lambda \rangle$
\begin{equation}\label{wflam}
W_{\lambda}(\mathbf{x},\mathbf{p})= \frac{1}{(2\pi \hbar)^4} \int
d\mathbf{u}\, d\mathbf{v} \langle \lambda | \exp{\big(
\frac{\emph{i}}{\hbar} \mathbf{u} \cdot
(\hat{\mathbf{p}}-\mathbf{p}) + \frac{ \emph{i}}{\hbar} \mathbf{v}
\cdot (\hat{ \mathbf{x}}- \mathbf{x}) \big)} | \lambda \rangle.
\end{equation}
Since the operators $\hat{x}$, $\hat{y}$, $\hat{p}_x$ and
$\hat{p}_y$ obey the commutation relations in Eq.(\ref{xp}), with the aid
of the Baker-Campbell-Hausdorff relation, one can write
\begin{eqnarray}
&&\langle \lambda |\exp{\frac{\emph{i}}{\hbar}(u_1 \hat{p}_x +u_2
\hat{p}_y + v_1 \hat{x} + v_2 \hat{y})} |\lambda \rangle
\nonumber\\
=&&\exp{\big(
\frac{\emph{i}}{2 \hbar^2}(\mu\,v_1 v_2 + \nu\,u_1 u_2)
+\frac{\emph{i}}{2 \hbar} (u_1 v_1 + u_2 v_2) \big)}
\nonumber\\
&&~~\times
\langle \lambda |\exp{(\frac{\emph{i}}{\hbar}v_1 \hat{x})}
\exp{(\frac{\emph{i}}{\hbar}u_1 \hat{p}_x )}
\exp{(\frac{\emph{i}}{\hbar}v_2
\hat{y})}\exp{(\frac{\emph{i}}{\hbar}u_2 \hat{p}_y )} |\lambda
\rangle .
\end{eqnarray}
Employing Eq.(\ref{lamxyp}) it is not difficult to evaluate the above matrix
element
\begin{multline}
\langle \lambda |\exp{(\frac{\emph{i}}{\hbar}v_1 \hat{x})}
\exp{(\frac{\emph{i}}{\hbar}u_1 \hat{p}_x )}
\exp{(\frac{\emph{i}}{\hbar}v_2
\hat{y})}\exp{(\frac{\emph{i}}{\hbar}u_2 \hat{p}_y )} |\lambda
\rangle
=  2\pi \hbar \delta (u_1 -u_2) \delta (v_1 +v_2) \\
\times\exp{\Big(\frac{\emph{i}}{4\hbar^2}\big( \mu (v_1^2 - 2v_1
v_2 - v_2^2) -\nu (u_1^2 +2 u_1 u_2 -u_2^2)\big)
-\frac{\emph{i}}{2\hbar}
(u_1 + u_2)(v_1 +v_2) \Big) } \\
\times\exp{\Big(\frac{\emph{i}}{\sqrt{2\hbar}} \big(
\sqrt[4]{\frac{\mu}{\nu}}(v_1 -v_2)(\lambda_1 - \theta \lambda_2) +
\sqrt[4]{\frac{\nu}{\mu}}(u_1 + u_2)(\lambda_2 - \theta \lambda_1)
\big) \Big)}.
\end{multline}
Substituting these results into Eq.(\ref{wflam}), after integrating over the
variables $\mathbf{u}$ and $\mathbf{v}$, one gets
\begin{equation}\label{wflam1}
W_{\lambda}(\mathbf{x},\mathbf{p})= \frac{1}{2\pi \hbar
\sqrt{1-\theta^2}} \delta \Big( x-y -\sqrt{2\hbar}
\sqrt[4]{\frac{\mu}{\nu}}(\lambda_1 - \theta \lambda_2 ) \Big)
\delta \Big( p_x + p_y -\sqrt{2\hbar}
\sqrt[4]{\frac{\nu}{\mu}}(\lambda_2 - \theta \lambda_1 ) \Big),
\end{equation}
which is consistent with the eigenvalue equations in Eq.(\ref{rplam}).

Furthermore, one can get an explicit form of
$\hat{\triangle}(\mathbf{x},\mathbf{p})$ in the entangled
representation $|\lambda \rangle$. Using Eq.(\ref{lamcpl}) one has
\begin{equation}
\hat{\triangle}(\mathbf{x},\mathbf{p})= \frac{1 - \theta ^2}{\pi^2
(2\pi \hbar)^4} \int d\mathbf{u} d\mathbf{v} d^2 \lambda d^2
\lambda' |\lambda \rangle \langle \lambda |  \exp{\big(
\frac{\emph{i}}{\hbar} \mathbf{u} \cdot
(\hat{\mathbf{p}}-\mathbf{p}) + \frac{ \emph{i}}{\hbar} \mathbf{v}
\cdot (\hat{ \mathbf{x}}- \mathbf{x}) \big)} |\lambda' \rangle
\langle \lambda' |.
\end{equation}
After integrating over the variables $\mathbf{u}$ and $\mathbf{v}$,
one gets
\begin{eqnarray}
\hat{\triangle}(\mathbf{x},\mathbf{p})&=& \frac{\sqrt{1-\theta^2}}{
2 \pi^3 \hbar} \int d^2 \lambda d^2 \lambda' \delta \Big( \lambda_1
+ \lambda_1^{'} - \theta (\lambda_2 + \lambda_2^{'})
-\sqrt{\frac{2}{\hbar}} \sqrt[4]{\frac{\nu}{\mu}} (x-y)
\Big)\nonumber\\
&\times&\delta \Big( \lambda_2 + \lambda_2^{'} - \theta (\lambda_1 +
\lambda_1^{'}) -\sqrt{\frac{2}{\hbar}} \sqrt[4]{\frac{\mu}{\nu}}
(p_x + p_y) \Big)
\nonumber\\
&\times&\exp{\Big(\emph{i}\,(\lambda_1 \lambda_2^{'} - \lambda_2
\lambda_1^{'}) +\emph{i}\,\frac{\theta}{2}(\lambda_1^{'2}-
\lambda_2^{'2}-\lambda_1^2 +\lambda_2^2)\Big)} \nonumber\\
&\times&\exp{ \Big(-\emph{i}\,
\sqrt{\frac{2}{\hbar}}\big(\sqrt[4]{\frac{\mu}{\nu}} p_y (\lambda_1
- \lambda_1^{'}) + \sqrt[4]{\frac{\nu}{\mu}} y (\lambda_2
-\lambda_2^{'}) \big) \Big)} |\lambda \rangle \langle \lambda' |.
\end{eqnarray}
Then performing the integration over $\lambda$ and introducing the
following notations
\begin{eqnarray}
\frac{1}{\sqrt{2\hbar}(1- \theta^2)}\sqrt[4]{\frac{\nu}{\mu}} (x-y)
=\rho_1, && ~~~\frac{1}{\sqrt{2\hbar}(1-
\theta^2)}\sqrt[4]{\frac{\mu}{\nu}} (p_x + p_y) =\rho_2, \nonumber\\
\frac{1}{\sqrt{2\hbar}(1- \theta^2)}\sqrt[4]{\frac{\nu}{\mu}} (x+y)
=\gamma_1, && ~~~\frac{1}{\sqrt{2\hbar}(1-
\theta^2)}\sqrt[4]{\frac{\mu}{\nu}} (p_x - p_y) =\gamma_2,
\end{eqnarray}
one can write the above expression as
\begin{multline}\label{delxp1}
\hat{\triangle}(\mathbf{x},\mathbf{p})=\\
=\frac{\sqrt{1-\theta^2}}{\pi^3
\hbar^2 } \int d^2 \lambda' \exp{\Big( -2\emph{i}\,\big( (\rho_2
-\theta \rho_1)(\lambda_1^{'} - \theta \lambda_2^{'}) -(\rho_1
-\theta \rho_2)(\lambda_2^{'} - \theta \lambda_1^{'}) \big)
\Big)} \\
~~~\times\exp{\Big( 2\emph{i}\,(1-\theta^2) \big( (\rho_2  -
\gamma_2) \lambda_1^{'} - (\rho_1  - \gamma_1) \lambda_2^{'} -
\gamma_1 (\rho_2 + \theta \rho_1) + \gamma_2 (\rho_1 + \theta
\rho_2)
\big)\Big)} \\
\times\big|2(\rho_1 + \theta \rho_2) -\lambda_1^{'}, 2(\rho_2 +
\theta \rho_1) -\lambda_2^{'} \big\rangle \big\langle \lambda_1^{'},
\lambda_2^{'}\big|~.
\end{multline}
At last, in Eq.(\ref{delxp1}) using new integration variables $\lambda$ to
replace the old ones $\lambda'$
\begin{equation}
\rho_1 + \theta \rho_2 - \lambda_1^{'} = \lambda_1, ~~~~~~~~ \rho_2
+ \theta \rho_1 - \lambda_2^{'} = \lambda_2,
\end{equation}
one obtains the expression of the Wigner operator in the entangled
state $|\lambda \rangle$ representation
\begin{eqnarray}\label{delxp2}
\lefteqn{\hat{\triangle}(\mathbf{x},\mathbf{p})=
\frac{\sqrt{1-\theta^2}}{\pi^3 \hbar^2 } \int d^2 \lambda \exp{\Big(
2\emph{i}\,(1-\theta^2) (\gamma_1 \lambda_2 -\gamma_2
\lambda_1)\Big)} }\nonumber\\
&&~~~~\times\big|\rho_1 + \theta \rho_2 - \lambda_1,
\rho_2 + \theta \rho_1 -
\lambda_2 \big\rangle \big\langle \rho_1 + \theta \rho_2 +
\lambda_1, \rho_2 + \theta \rho_1 + \lambda_2 \big|,
\end{eqnarray}
where the ket and bra vector are in the $ |\lambda \rangle$
representation. Since the entangle states $|\lambda \rangle$ are
orthonormal to each other (see Eq.(\ref{lamorth1})), it is very easy to get the
WF $W_\lambda (\mathbf{x},\mathbf{p})$ (\ref{wflam1}) from the expression
(\ref{delxp2}). The only needed to do is take the expectation value of the
Wigner operator (\ref{delxp2}) in the state $|\lambda \rangle$.

Similarly, one can get the WF of the entangled state $|\xi \rangle$
\begin{equation}
W_{\xi}(\mathbf{x},\mathbf{p})= \frac{1}{2\pi \hbar
\sqrt{1-\theta^2}} \delta \Big( x+y -\sqrt{2\hbar}
\sqrt[4]{\frac{\mu}{\nu}}(\xi_1 + \theta \xi_2 ) \Big) \delta \Big(
p_x - p_y -\sqrt{2\hbar} \sqrt[4]{\frac{\nu}{\mu}}(\xi_2 + \theta
\xi_1 ) \Big),
\end{equation}
which is consistent with the eigenvalue equations in Eq.(\ref{qkxi}). Using
the inner product (\ref{lamxi}), from Eq.(\ref{delxp2}) one has the explicit expression
of the Wigner operator in the entangled state $|\xi \rangle$
representation
\begin{eqnarray}
\lefteqn{\hat{\triangle}(\mathbf{x},\mathbf{p})=
\frac{\sqrt{1-\theta^2}}{\pi^3 \hbar^2 } \int d^2 \xi \exp{\Big(
2\emph{i}\,(1-\theta^2) (\rho_1 \xi_2 -\rho_2 \xi_1)\Big)}}\nonumber\\
&&~~~~\times\big|\gamma_1 - \theta \gamma_2 - \xi_1, \gamma_2 - \theta
\gamma_1 - \xi_2 \big\rangle \big\langle \gamma_1 - \theta \gamma_2
+ \xi_1, \gamma_2 - \theta \gamma_1 + \xi_2 \big|,
\end{eqnarray}
where the ket and bra vector are in the $ |\xi \rangle$
representation.

\section{Some possible applications}\label{sec4}
It is well know that representation plays a basic role in quantum
mechanics like the coordinate systems in geometry. In section \ref{sec2} we
introduced the entangled state representations $|\lambda \rangle$
and $|\xi \rangle$. In the $|\lambda \rangle$ or $|\xi \rangle$
representation one can also solve problems of NCQM, and sometimes it
is more convenient working in the entangled state representation
than in other representation. To show this, let us study a
two-dimensional harmonic oscillator on the noncommutative phase
space with both momentum-momentum (kinetic) coupling and
coordinate-coordinate (elastic) coupling. The quantum Hamiltonian is
\begin{equation}\label{ham}
\hat{H}=\frac{1}{2m}\left(\hat{p}_x^{2}+
\hat{p}_y^{2} \right) +\frac{m\omega^2}{2} \left( \hat{x}^{2}+
\hat{y}^{2} \right) + \frac{k}{2} \left( \hat{x}\hat{y} + \hat{y}
\hat{x} \right) + \frac{l}{2} \left( \hat{p}_x \hat{p}_y + \hat{p}_y
\hat{p}_x \right),
\end{equation}
where the operators $\hat{p}_{x}$, $\hat{p}_{y}$, $\hat{x}$ and
$\hat{y}$ satisfy the commutation relations (\ref{xp}). If one use the
usual Seiberg-Witten map to rewrite this Hamiltonian in terms of
some kind of commutative coordinate and momentum operators, one will
have a very complicated expression, which includes not only the
kinetic and the elastic coupling terms, but also the
coordinate-momentum coupling terms (they are the angular momentum
term and the squeezing term, respectively). It is not an easy task
to solve its energy spectra. However, in the $|\lambda \rangle$
representation the Hamiltonian $\hat{H}$ (\ref{ham}) has a very simple form
\begin{equation}\label{ham1}
\hat{H}=c_1 \eta_{1}^2 + c_2 \eta_{2}^2 + d_1 \lambda_{1}^2 + d_2
\lambda_{2}^2 + d_3 \lambda _1 \lambda_2~,
\end{equation}
where $\eta_1 =-\emph{i}\,\partial /\partial \lambda_1$, $\eta_2
=-\emph{i}\,\partial /\partial \lambda_2$, which satisfy the
standard Heisenberg commutation relations
\begin{equation}
[\lambda_i,\,\eta_j]=\emph{i}\,\delta_{ij}\,, ~~~~~~~~
[\lambda_1,\,\lambda_2]=[\eta_1,\,\eta_2]=0.
\end{equation}
In Eq.(\ref{ham1}), the coefficients $c_1,\,c_2,\,d_1,\,d_2,\,d_3$ are
\begin{eqnarray}
&&c_1=\frac{\hbar}{2m}\sqrt{\frac{\nu}{\mu}}\left( 1-l m
\right),~~~~~~ c_2=\frac{\hbar m \omega^2}{2} \sqrt{\frac{\mu}{\nu}}
\left( 1+ \frac{k}{m \omega^2}\right), \nonumber\\
&&d_1=\frac{\hbar \theta^2}{2m} \sqrt{\frac{\nu}{\mu}}\left(1 +l m
\right) + \frac{\hbar m \omega^2}{2} \sqrt{\frac{\mu}{\nu}} \left(
1- \frac{k}{m \omega^2} \right), \nonumber\\
&&d_2=\frac{\hbar}{2m} \sqrt{\frac{\nu}{\mu}}(1 +l m) + \frac{\hbar
\theta^2 m \omega^2}{2}
\sqrt{\frac{\mu}{\nu}} \left( 1- \frac{k}{m \omega^2} \right), \nonumber\\
&&d_3=-\frac{\nu}{m} \left( 1+l m \right) -\mu m \omega^2 \left( 1 -
\frac{k}{m \omega^2} \right),
\end{eqnarray}
respectively. Thus the two-dimensional harmonic oscillator on the
noncommutative phase space with both momentum-momentum (kinetic)
coupling and coordinate-coordinate (elastic) coupling is transferred
to a very simple ordinary two-dimensional coupling oscillator in the
$|\lambda \rangle$ representation, and the corresponding energy
spectra and WFs are easily to get.

In fact, one can make a scaling change and a planar rotation to
reduce the Hamiltonian (\ref{ham1}) further \cite{s15,s16}. To do this, we make
the following scaling changes
\begin{equation}\label{lameta}
\lambda_1'=\sqrt[4]{\frac{c_2}{c_1}}\,\lambda_1\,,~~~~
\lambda_2'=\sqrt[4]{\frac{c_1}{c_2}}\,\lambda_2\,,~~~~
\eta_1'=\sqrt[4]{\frac{c_1}{c_2}}\,\eta_1\,,~~~~
\eta_2'=\sqrt[4]{\frac{c_2}{c_1}}\,\eta_2\,,
\end{equation}
which lead the Hamiltonian (\ref{ham1}) to
\begin{equation}\label{ham2}
\hat{H}=c\left( \eta_1'^2 +\eta_2'^2 \right) + f_1 \lambda_1'^2 +
f_2 \lambda_2'^2 +d_3 \lambda_1' \lambda_2'~,
\end{equation}
where
\begin{eqnarray}
c&=&\sqrt{c_1 c_2}=\frac{\hbar \omega}{2} \sqrt{\left(1-l m \right)
\left( 1+ \frac{k}{m \omega^2} \right)}~, \nonumber\\
f_1 &=&d_1 \sqrt{\frac{c_1}{c_2}}
=\left( \frac{\hbar \omega}{2} \left(1-
\frac{k}{m \omega^2} \right) + \frac{\nu^2}{2m^2 \hbar
\omega}\left(1 + l m \right) \right) \sqrt{\frac{1-l m}{1+
\frac{k}{m \omega^2}}}~, \nonumber\\
f_2 &=&d_2  \sqrt{\frac{c_2}{c_1}}
=\left( \frac{\hbar \omega}{2} \left(1+ l
m \right) + \frac{\mu^2 m^2 \omega^3}{2 \hbar}\left(1 - \frac{k}{m
\omega^2} \right) \right) \sqrt{\frac{1+ \frac{k}{m \omega^2}}{1- l
m}}~.
\end{eqnarray}
Then, since the Hamiltonian (\ref{ham2}) still involves an interaction term,
we can simplify this situation by a transformation to new phase
space variables
\begin{equation}\label{xipi}
x_i = u_{i j}\lambda_j',~~~~~~~~p_i=u_{i j}\eta_j',~~~~~~~~(i,j=1,2)
\end{equation}
where the matrix
\begin{equation}
(u_{i j})=\begin{pmatrix}
\cos \frac{\alpha}{2} & -\sin\frac{\alpha}{2} \\
\sin \frac{\alpha}{2} & \cos \frac{\alpha}{2} \\
\end{pmatrix}
\end{equation}
is a unitary rotation with the mixing angle $\alpha$. When $\alpha$
satisfies the condition
\begin{equation}
\tan \alpha =\frac{d_3}{f_2 - f_1},
\end{equation}
the Hamiltonian (\ref{ham2}) has a factorizing form
\begin{equation}\label{ham3}
\hat{H} =\frac{1}{2M}p_1^2 + \frac{M \Omega_{+}^2}{2}x_1^2 +
\frac{1}{2M}p_2^2 + \frac{M \Omega_{-}^2}{2}x_2^2 \,,
\end{equation}
where
\begin{eqnarray}
M&=&\left( \hbar \omega \sqrt{\left(1- l m \right) \left( 1+
\frac{k}{m \omega^2} \right)}~ \right)^{-1}, \nonumber\\
\Omega_{\pm}&=&\hbar\omega \left(\sqrt{ A_{+} + B_{-}}\,
\pm \sqrt{ A_{-} + B_{+}}~\right),
\end{eqnarray}
where
\begin{eqnarray}
A_{\pm}&=&\frac{1}{2}\left(1+ \frac{k
l}{\omega^2} \pm \sqrt{\left(1- l^2 m^2 \right)\Big(1 -
\frac{k^2}{m^2 \omega^4} \Big) }~\right),\nonumber\\
B_{\pm}&=&\left( \frac{\nu
\sqrt{1- l^2 m^2}}{2m\hbar\omega} \pm \frac{\mu m \omega}{2\hbar} \sqrt{1-
\frac{k^2}{m^2
\omega^4}}~\right)^2.\nonumber
\end{eqnarray}
Obviously, the transformations (\ref{lameta}) and (\ref{xipi}) do not change the
standard Heisenberg commutation relations, so the Hamiltonian (\ref{ham3})
describes a two-dimensional uncoupling harmonic oscillator and its
energy level as well as WFs are well-known
\begin{equation}\label{en}
E_{n_1 n_2}=(n_1 +\frac{1}{2})\Omega_{+} + (n_2 +
\frac{1}{2})\Omega_{-},
\end{equation}
and
\begin{equation}
W_{n_1 n_2}=4(-1)^{n_1 +n_2} e^{-2H_{1}/\Omega_{+}}
e^{-2H_{2}/\Omega_{-}} L_{n_1}\left( \frac{4H_{1}}{\Omega_{+}}
\right) L_{n_2}\left( \frac{4H_{2}}{\Omega_{-}} \right)
\end{equation}
respectively, where $L_n$ are the Laguerre polynomials and $H_1
=\frac{1}{2M}p_1^2 + \frac{M \Omega_{+}^2}{2}x_1^2 $\,, $H_2
=\frac{1}{2M}p_2^2 + \frac{M \Omega_{-}^2}{2}x_2^2$\,, and here $x_1,
\,x_2$ and $p_1,\,p_2$ are classical phase space variables.

The energy (\ref{en}) will reduce to some familiar results in several
specific situations, for instance, when $\mu,\nu=0$, i.e., in the
ordinary commutative phase space, (\ref{en}) becomes
\begin{eqnarray}
E_{n_1 n_2}&=&(n_1 + n_2 +1) \frac{\hbar \omega}{\sqrt{2}} \sqrt{1+
\frac{k l}{\omega^2}+ \sqrt{\left(1- l^2 m^2 \right) \left(1-
\frac{k^2}{m^2
\omega^4} \right)}} \nonumber\\
&&+ ( n_1 -n_2) \frac{\hbar \omega}{\sqrt{2}} \sqrt{1+ \frac{k
l}{\omega^2} - \sqrt{\left(1- l^2 m^2 \right) \left(1-
\frac{k^2}{m^2 \omega^4} \right)}}~;
\end{eqnarray}
when $k,l=0$ and $\mu,\nu \neq 0$, (\ref{en}) becomes
\begin{equation}
E_{n_1 n_2}=(n_1 +n_2 +1) \hbar \omega \sqrt{1+ \frac{(\nu- \mu m^2
\omega^2)^2}{4m^2 \hbar^2 \omega^2}}\, + (n_1 -n_2) \left(
\frac{\nu}{2m} + \frac{\mu m \omega^2}{2} \right);
\end{equation}
and when both $\mu,\nu=0$ and $k,l=0$, the energy has familiar form
$E_{n_1 n_2}= (n_1 +n_2 +1) \hbar \omega$.

\section{Summary and discussion}\label{sec5}
In order to develop representation and transformation theory so that
one can solve more dynamic problems for NCQM, employing the EPR idea
we construct simultaneous eigenstates of the difference (or the sum)
of two different coordinate-component operators and the sum (or the
difference) of two relevant momentum operators. Since these new
state-vectors are orthonormal and satisfy the completeness relation,
they form representations to formulate the NCQM and we name them the
entangled state representations. We also derive explicit expressions
of Wigner operator and WFs in the new representations. In order to
show the potential role of the new representations in NCQM, we study
a two-dimensional oscillator with both kinetic and elastic couplings
on the noncommutative phase space and simply get its exact energy
spectra and WFs in the entangled state representation.

It is worth pointing out that the new entangled state
representations are built based on the deformed boson algebra (\ref{abcr}).
Also form the deformed boson algebra (\ref{abcr}), one can construct the
coherent state representation and the squeezed state representation
on the noncommutative phase space. Thus although there
are no the standard coordinate and the standard momentum
representations on the noncommutative phase space, one can develop
the coherent state, the squeezed state and the entangled state and
other useful representations to solve physics problems on the
noncommutative phase space. We would like also to emphasize that, in
contrast to the most work on the NCQM in the literature, our work is
done directly in the noncommutative phase space without using any
variables in the ordinary commutative phase space. So if one thinks
the usual way to NCQM (transferring problems in the noncommutative
space into corresponding ones in the commutative space) as the
``perturbation approach'', our way may be named as ``nonperturbation
approach''.

It is also interesting to develop the theory of representations and
transformations on the noncommutative phase space further and work
on this direction will be presented in a separate paper.

\section*{Acknowledgments}
This project was supported by the National Natural Science Foundation of China
under Grant 10675106.

\end{document}